\documentclass[aps,pra,nofootinbib,twocolumn,floatfix,showpacs,superscriptaddress]{revtex4-2}
\usepackage{amsmath,graphicx,epsfig,amsthm,color,bm}
\usepackage{pifont,bbding,amssymb,soul,mathrsfs,braket}

\newcommand{\red}[1]{{\color{red} #1}}

\usepackage{threeparttable}
\usepackage{multirow}
\usepackage[hyperindex,breaklinks,colorlinks=true,citecolor=blue,urlcolor=blue]{hyperref}
\usepackage{appendix, comment}

\begin{document}

\title{Compact non-degenerate entangled-photon source and near-infrared-to-telecom quantum teleportation}
\author{Xu-Jie Peng}
\email{These authors contributed equally to this work.}
\affiliation{School of Physics, State Key Laboratory of Crystal Materials, Shandong University, Jinan 250100, China.}
\affiliation{Shenzhen Research Institute of Shandong University, Shenzhen 518057, China}

\author{Ling-Xuan Kong}
\email{These authors contributed equally to this work.}
\affiliation{School of Physics, State Key Laboratory of Crystal Materials, Shandong University, Jinan 250100, China.}
\affiliation{Shenzhen Research Institute of Shandong University, Shenzhen 518057, China}

\author{He Lu}
\email{luhe@sdu.edu.cn}
\affiliation{School of Physics, State Key Laboratory of Crystal Materials, Shandong University, Jinan 250100, China.}
\affiliation{Shenzhen Research Institute of Shandong University, Shenzhen 518057, China}

\begin{abstract}
The polarization-entangled photon source (PEPS) at non-degenerated wavelengths is pivotal to connect quantum systems working at different wavelengths, with the assistance of quantum teleportation. Here, a compact Sagnac-type photon source is designed and demonstrated, in which two photons with wavelengths at 810 and 1550~nm are highly entangled in polarization degree of freedom. The two photons are generated from a periodically poled lithium niobate crystal pumped with a 532~nm continuous-wave laser, via type-0 nondegenerate spontaneous parametric down-conversion. The polarization of three lights is rotated by a single periscope, which makes the Sagnac interferometer compact and stable. The generated two photons are with high brightness of $3\times10^4$ pairs/s/mW, which are highly entangled with fidelity of $0.985\pm0.002$. The entanglement is verified by violating the Clauser-Horne-Shimony-Holt inequality with $\mathcal S =2.756\pm0.007$. Finally, teleportation is demonstrated with this nondegenerate source, in which photonic states at 810~nm is teleported to 1550~nm with fidelity of $0.955\pm0.003$.
\end{abstract}

\maketitle

Spontaneous parametric down-conversion~(SPDC) has been developed as a mature technique to generate photon pairs~\cite{Burnham1970PRL}, which is further tailored to entangled in various degree of freedom~(DoF) of photons, including polarization~\cite{Kwiat1995PRL}, spatial modes~\cite{Walborn2010PR,Edgar2012NC}, time-bin~\cite{Jin2024PQE} and frequency~\cite{Kuzucu2005PRL}.  Notably, polarization-entangled photon source ~(PEPS) from degenerate SPDC, i.e., the two photons are with identical wavelength, has been widely adopted in quantum teleportation~\cite{Bouwmeester1997Nature,Ma2012Nature,Yin2012Nature,Ren2017Nature}, which is the crucial mechanism for building large-scale quantum network~\cite{Kimble2008Science}. The quantum network has emerged as an advanced quantum technology as it enables the connection of quantum systems over long distances, thus greatly enhancing the capability of information processing. Remarkable achievements have extended the distance to hundreds of kilometers with fibre links~\cite{Martin2021EPJ}, and thousands of kilometers with free space links~\cite{Yin2017Science,Ren2017Nature}. The operating wavelength(s) of the quantum network is complex and depends on the wavelength of quantum device at each node. The favorable choice is near-infrared~(NIR) and C-band wavelengths~\cite{Parny2023CP}, as the NIR wavelengths are compatible with free-space links and atmospheric transmission windows~\cite{Yin2012Nature,Steinlechner2017NC,Yin2017Science,Ren2017Nature} while C-band wavelength is compatible with fiber devices and ground quantum network~\cite{Yin2016PRL,Chen2020PRL}. 

Along this spirit, tremendous efforts have been devoted to develop non-degenerate PEPS with vastly separated wavelengths, which are capable of bridging different transmission channels within a quantum network. There are mainly two design drivers to implement a high-quality nondegenerate PEPS. The first is the choice of non-linear material as the non-linearity determines the brightness of PEPS. Periodically poled potassium titanyl phosphate~(PPKTP)~\cite{Pelton2004OE,Fiorentino2008OE,Hentschel2009OE,Chen2009OE,Clausen2014NJP,Aizawa2023AO,Hong2024OL,Fallon2025OE} and periodically poled lithium niobate~(PPLN)~\cite{Konig2005PRA,deChatellus2006OE,Hubel2007OE,Sauge2008OE,Stuart2013PRA,Clausen2014NJP,Dietz2016APB,Szlachetka2023APL} are most commonly used. Polarization maintaining fiber~(PMF) is able to generate PEPS via spontaneous four-wave mixing~(SFWM)~\cite{Lee2021QST}, however, the third-order nonlinearity~($\chi^{(3)}$) is SFWM is much weaker than second-order nonlinearity~($\chi^{(2)}$) in SPDC. Compared to KTP, LN has larger$\chi^{(2)}$ that admits higher brightness of PEPS. The second issue is the geometry of PEPS that can be summarized into two categories: single crystal-double pump~(SC-DP)~\cite{Konig2005PRA,Sauge2008OE,Fiorentino2008OE,Hentschel2009OE,Dietz2016APB,Aizawa2023AO,Lee2021QST,Szlachetka2023APL,Hong2024OL} and double crystals-single pump~(DC-SP)~\cite{Pelton2004OE,Hubel2007OE}. In DC-SP, a single light pumping two nonlinear crystals is used, and the photons from the two SPDC are combined interferometrically. It is crucial that the two SPDC sources be identical, as the source difference would introduce distinguishability and thus decrease the fidelity of PEPS~\cite{Fiorentino2004PRA}. This can be addressed by SC-DP schemes as photons are generated from the same crystal, which can be realized by Mach-Zehnder interferometer~(MZI)~\cite{Konig2005PRA,Sauge2008OE,Fiorentino2008OE} , Sagnac interferometer~\cite{Hentschel2009OE,Aizawa2023AO}, folded-sandwich configuration~\cite{Dietz2016APB} and MZI+Sagnac~\cite{Lee2021QST,Szlachetka2023APL,Hong2024OL}. Compared to MZI, Sagnac interferometer and folded-sandwich configuration  admit intrinsic phase stability, which benefits applications requiring long-term-phase-stable PEPS~\cite{Kim2006PRA,Shi2004PRA}. However, folded-sandwich configuration requires double length crystal to compensate the dispersion. One of the challenges in building up a Sagnac-type PEPS is to rotate the polarization for all three wavelengths. The conventional periscope consisting of two reflective mirrors works for all wavelength~\cite{Szlachetka2023APL}, yet suffering alignment difficulties. 
\begin{figure*}[h!tbp]
\centering
\includegraphics[width=\linewidth]{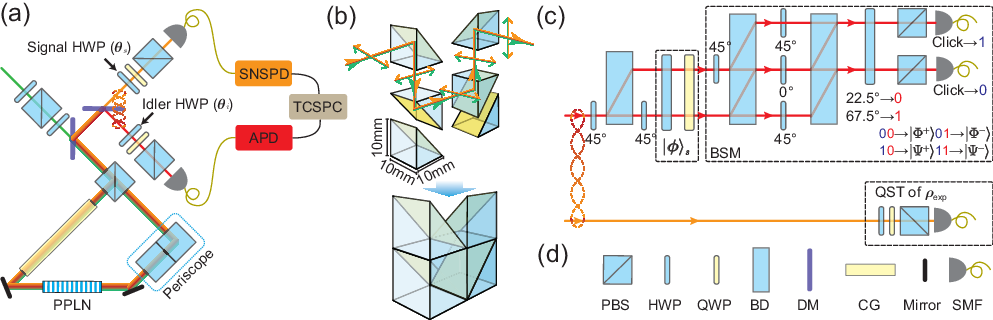}
\caption{\label{Fig:setup}The schematic drawing of experimental setup. a) The setup of Sagnac-type nondegenerate PEPS. b) The compact periscope. The  blue-marked surface corresponds to anti-reflection coating, while the one yellow-marked surface corresponds to reflection coating. The six prisms are bonded together using a high-transmittance optical adhesive. c) The setup to perform teleportation. d) Symbols used in (a) and (c). PBS: polarizing beam splitter; HWP: half-wave plate; QWP: quarter-wave plate; BD: beam displacer; DM: dichroic mirror; CG: compensation glass; SMF: single-mode fiber.}
\end{figure*}

In this work, we tackle this issue with a novel periscope consisting of six right-angle prisms, which enables polarization rotation for broadband wavelengths. Equipped with this periscope, we demonstrate Sagnac-type nondegenerate PEPS with wavelengths at 810 and 1550~nm, generated by a PPLN pumped by 532~nm continuous-wave~(CW) laser via type-0 SPDC. The measured coincidence is $3\times10^4$ pairs/s/mW, and the photon generation rate~(PGR) is $8.2\times10^5$ pairs/s/mW. The entanglement is verified by fidelity with respect to maximally entangled of $0.985\pm0.002$, and violation of the Clauser–Horne–Shimony–Holt~(CHSH) inequality with $\mathcal S =2.756\pm0.007$. The application of our source is demonstrated in a teleportation experiment with fidelity of $0.955\pm0.003$. 
\begin{figure*}[ht!bp]
\includegraphics[width=\linewidth]{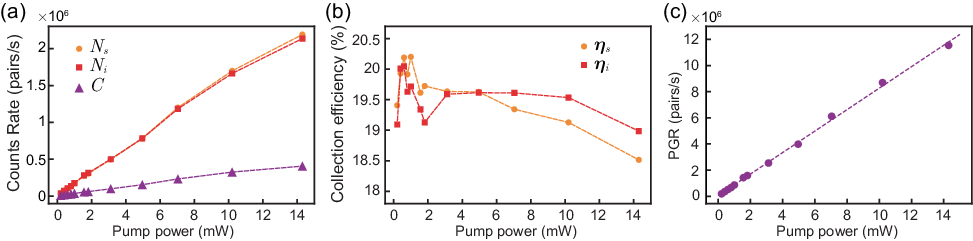}
\caption{\label{Fig:brightness} a) The counts rate of signal photon $N_s$~(orange dots), idler photon $N_i$~(red squares) and two-photon coincidence $C$~(purple triangles) at different pump powers. b) The collection efficiency of signal photon $\eta_s$~(orange dots) and idler photon $\eta_i$~(red squares) at different pump powers. c) The photon pair generation rate~(PGR) at different pump powers.  All the data are collected in the computational basis. The statistic errors are too small compared to the size of data points.}
\end{figure*}

The PPLN crystal used in our experiment is with size of $1\times1\times10$ mm$^3$. The poling period is 7.1~$\mu$m, satisfying the quasi-phase matching in type-0 nondegenerate SPDC of $532~nm\to810~\text{nm}+1550~\text{nm}$, where the signal and idler photons have the same polarization as pump light, i.e., $o\to o+o$. As shown in Figure~\ref{Fig:setup}~a, we use a 532~nm continuous-wave~(CW) laser to coherently pump the PPLN crystal from the clockwise and counterclockwise directions. The power of the pump light can be adjusted by a half-wave plate~(HWP) and a polarizing beam splitter~(PBS), where PBS transmits the horizontal polarization~($\ket{H}$) and reflects the vertical polarization~($\ket{V}$). The horizontal polarization $\ket{H}_p$ is then rotated to $\ket{D}_p=\frac{1}{\sqrt{2}}(\ket{H}_p+\ket{V}_p)$ by a HWP set at 22.5$^\circ$. The second PBS coherently splits the polarizations of pump light, i.e., $\ket{H}_p$ and $\ket{V}_p$ propagate along clockwise and counterclockwise directions respectively. Pump beam is focused into PPLN crystal with beam waist of 130~$\mu$m by two plano-convex lenses.

The structure of our periscope is illustrated in Figure~\ref{Fig:setup}~b. It comprises six right-angle prisms made of K9 glass with a side length of 10~mm. The two blue-marked surfaces correspond to anti-reflection coating, while the one yellow-marked surface correspond to reflection coating. Light incident with horizontal or vertical polarization undergoes four reflections via four prisms. The fourth prism is twisted by 90$^\circ$, which equivalently converts the polarization of output light to vertical or horizontal polarization respectively. The six prisms are bonded together using a high-transmittance optical adhesive.  Note that the prism admits reflection of wavelengths from 532 to 1550~nm, so that it is capable to implement polarization rotation of wavelengths involved in this experiment. The measured transmission efficiency of the periscope is 94.6\% at 532~nm, 94.4\% at 810~nm, and 94.0\% at 1550~nm, respectively. The optical loss induced by the periscope slightly reduces the brightness of PEPS. With such a novel periscope, the $\ket{H}_p$ component, propagating in the clockwise direction, is converted to $\ket{V}_p$. Consequently, the clockwise SPDC generates $\ket{V}_s\ket{V}_i$ in PPLN. In counterclockwise SPDC, the component $\ket{V}_p$ generates $\ket{V}_s\ket{V}_i$ in PPLN, which is then converted to $\ket{H}_s\ket{H}_i$ by the periscope. Finally, $\ket{V}_s\ket{V}_i$ and $\ket{H}_s\ket{H}_i$ is combined on PBS, leading to the output state in the form of 
\begin{equation}\label{Eq:MES}
\ket{\Phi^+}=\frac{1}{\sqrt{2}}\left(\ket{H}_s\ket{H}_i+\ket{V}_s\ket{V}_i\right), 
\end{equation}
where the wavelengths of signal and idler photons are 1550 and 810~nm respectively. The periscope introduces time delay between the photons generated under clockwise-pumped
and counterclockwise-pumped conditions, thus affecting the interference on PBS. To compensate this delay, a compensation glass~(CS) made of K9 glass is placed in the Sagnac interferometer. The length of the CS is selected to be 40~mm, matching the optical path length that photons traverse in the periscope. The signal and idler photons are separated by two dichroic mirrors~(DMs), which are then coupled into single-mode fibre for detection. We use a superconducting nanowire single-photon detector~(SNSPD, detection efficiency $\approx$90\%) and a silicon avalanche photodetector~(APD, detection efficiency $\approx$60\%) to detect signal and idler photons respectively, and record the data with a time-correlated single photon counting~(TCSPC) system. The projective measurement of signal~(idler) photon on arbitrary state $\ket{\phi_{s(i)}}$ is implemented by a HWP, a quarter-wave plate~(QWP) and a PBS.

We first measure the count rate of signal photon~($N_s$), idler photon~($N_i$)  and two-photon coincidence~($C$) with pump power of 0.2, 0.4, 0.6, 0.8, 1.0, 1.6, 1.8, 3.1, 5.0, 7.0, 10.2 and 14.3~mW, respectively. $N_s$, $N_i$ and $C$ are measured in the computational basis, i.e., 
\begin{equation}
\begin{split}
&N_s=N_{H_s}+N_{V_s},\\
&N_i=N_{H_i}+N_{V_i},\\
&C=N_{H_sH_i}+N_{H_sV_i}+N_{V_sH_i}+N_{V_sV_i},
\end{split}
\end{equation}
and the results are shown in Figure~\ref{Fig:brightness}~a. By linearly fitting the coincidence $C$ with different pump power, we obtain the brightness of $3\times10^4$~pairs/s/mW. Triggering signal~(idler) photon heralds the existence of the idler~(signal) photon, which is known as heralded single photon source~(HSPS). The heralding
efficiency $\eta_{i(s)}$ of HSPS is determined $\eta_{i(s)}=C/N_{s(i)}$, which is shown in Figure~\ref{Fig:brightness}~b. The heralding efficiency is greater than 19\% with pump power is less than 10~mW. At high pump power, heralding efficiency decreases slightly, which is caused by the multiphoton excitation in SPDC. The pair generation rate~(PGR) is defined as 
\begin{equation}
\text{PGR}=\frac{N_s\times N_i}{C},
\end{equation}    
and the results are shown in Figure~\ref{Fig:brightness}~c. The PGR obtained by linear fitting is $8.2\times10^5$~pairs/s/mW. 

 \begin{figure}[h!tbp]
\includegraphics[width=\linewidth]{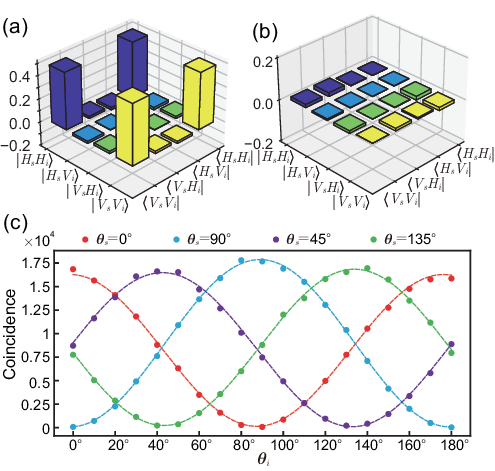}
\caption{\label{Fig:entanglement} a) Real part of the reconstructed $\rho_\text{exp}$. b) Imaginary part of the reconstructed $\rho_\text{exp}$. c) Polarization correlation between signal and idler photons. The red, blue, purple and green dots represent the results by fixing $\theta_s$ at $0^\circ$, $45^\circ$, $90^\circ$ and $135^\circ$, respectively.} 
\end{figure}

The entanglement is first characterized by fidelity of prepared $\rho_\text{exp}$ with respect to the ideal form in Equation~\ref{Eq:MES}, i.e., $F=\text{Tr}(\rho\ket{\Phi^+}\bra{\Phi^+})$. To this end, we perform quantum state tomography on $\rho_\text{exp}$~\cite{James2001PRA}, and the reconstructed $\rho_\text{exp}$ are shown in Figure~\ref{Fig:entanglement}~a,b. The data are collected with pump power of 1~mW, where the brightness is $3.6\times 10^4$~pairs/s. According to the reconstructed $\rho_\text{exp}$, we calculate the fidelity to be  $F=0.985\pm0.002$. The entanglement is also verified by the polarization correlation between signal and idler photons. We fix the angle of HWP set in signal path and record the coincidence at different angles of HWP set in idler path. As shown in Figure~\ref{Fig:entanglement}~c, the polarization correlation is measured by setting the angle of signal HWP $\theta_s=0^\circ, 45^\circ, 90^\circ$ and $135^\circ$ respectively. For each fixed $\theta_s$, the angle of idler HWP $\theta_i$ is set from $0^\circ$ to $180^\circ$ with interval of $10^\circ$. The data are sinusoidally fitted, according to which we calculate the visibilities of $\mathcal V(0^\circ)=98.95\%$, $\mathcal V(45^\circ)=99.46\%$, $\mathcal V(90^\circ)=99.69\%$ and $\mathcal V(135^\circ)=97.95\%$, where the visibility is defined by 
\begin{equation}
\mathcal V(\theta_s)=\frac{C_\text{max}-C_\text{min}}{C_\text{max}+C_\text{min}}
\end{equation} 
with $C_\text{max}$ and $C_\text{min}$ being the maximal and minimum coincidence. We also test the CHSH inequality on $\rho_\text{exp}$, in form of~\cite{Clauser1969PRL} 
\begin{equation}
\mathcal S=P(\theta_s, \theta_i)+P(\theta_s^\prime, \theta_i^\prime)+P(\theta_s, \theta_i^\prime)-P(\theta_s^\prime, \theta_i). 
\end{equation}
$P(\theta_s, \theta_i)$ is the probability of projecting $\rho_\text{exp}$ on $\ket{\psi_{\theta_s}}\ket{\psi_{\theta_i}}$ with 
\begin{equation}
\ket{\psi_{\theta_{s(i)}}}=\cos\frac{\theta_{s(i)}}{2}\ket{H}_{s(i)}+\sin\frac{\theta_{s(i)}}{2}\ket{V}_{s(i)},
\end{equation} 
and $\theta_s$ and $\theta_i$ are the angles of signal HWP and idler HWP respectively. We observe $\mathcal S =2.756\pm0.007$, which violates the classical threshold $\mathcal S=2$ by 108 standard deviations.

\begin{figure}[htbp]
\includegraphics[width=\linewidth]{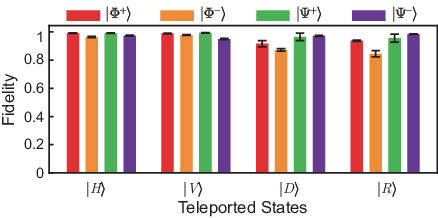}
\caption{\label{Fig:tele} Experimental results of state fidelities after teleportation. The red, orange, green and purple bars represents the fidelity with BSM outcomes of $\ket{\Phi^+}$, $\ket{\Phi^-}$, $\ket{\Psi^+}$ and \red{$\ket{\Psi^-}$}, respectively.} 
\end{figure} 

Using the prepared $\rho_\text{exp}$, we implement the two-photon teleportation
scheme ~\cite{Boschi1998PRL,Jin2010NP,Jiang2020SR,Li2021PRR,Zhang2024PRL} as illustrated in Figure~\ref{Fig:setup}~c, where a polarization encoded state $\ket{\phi}_{s}$ at 810~nm is teleported to state at 1550~nm. 
To this end, entanglement between signal and idler photon, encoded in polarization DoF, is first mapped to polarization DoF of signal photon and path DoF of idler photon using a beam displacer. Using a HWP and a QWP, we prepare select four states $\ket{\phi}_{s}$ to be teleported, i.e., $\ket{\phi}_{s}\in\{\ket{H}, \ket{V}, \ket{D}=\frac{1}{\sqrt{2}}(\ket{H}+\ket{V}), \ket{R}=\frac{1}{\sqrt{2}}(\ket{H}+i\ket{V})\}$. The Bell-state measurement~(BSM) is taken place between path DoF and polarization DoF of idler photon, and the output port along with the angle of HWP indicates the projection on four Bell states~(the dashed box ``BSM" in Figure~\ref{Fig:setup}~c). After teleportation, we perform the QST on signal photon to reconstruct the teleported state $\rho_\phi$, and then calculated the fidelity with respect to ideal case, i.e., $F=\text{Tr}(\rho_\phi U\ket{\phi}_{s}\bra{\phi}_{s}U^\dagger)$ with $U$ being the correction unitary depending on the outcome of BSM. Specifically, the outcome of $\ket{\Phi^+}$, $\ket{\Phi^-}$, $\ket{\Psi^+}$ and $\ket{\Psi^+}$ correspond to $U=XZ, X, Z, I$. The results of fidelities are shown in Figure~\ref{Fig:tele}, and we calculate the average fidelity of $\ket{H}, \ket{V}, \ket{D}$ and $\ket{R}$ after teleportation is $0.955\pm0.003$. The discrepancy between the experimental results and the ideal value is primarily attributed to experimental imperfections. Specifically, the higher-order emission of SPDC process decreases the quality of the entangled photons, while the spatial mismatch between the two interfering light beams degrades the interference quality at the BDs.

\begin{figure}[ht!bp]
    \centering
    \includegraphics[width=1\linewidth]{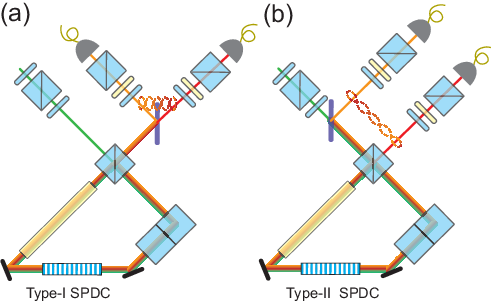}
    \caption{The schematic drawing of experimental setups to generate non-degenerate entangled photon sources with (a) type-I and (b) type-II SPDC. }
    \label{fig:type12}
\end{figure}

In conclusion, we design and demonstrate a compact Sagnac-type PEPS with nondegenerate wavelengths, i.e., signal and idler photons with 1550~nm and 810~nm. Compared to other Sagnac-type PEPS~\cite{Hentschel2009OE, Stuart2013PRA, Aizawa2023AO, Lee2021QST, Szlachetka2023APL}, the compact periscope used in our experiment alleviates the alignment difficulties, resulting PEPS with high fidelity and high heralding efficiency. We summarize several features of  non-degenerated PEPSs in Table~\ref{Tb:spdc}, including material, phase-matching type, PGR, coincidence, wavelengths, source setting, fidelity and heralding efficiency. Our source exhibits both a high entanglement fidelity of 98.5\% and a heralding efficiency of 19.1\%---performance metrics that are competitive with those of most previously reported non-degenerate PEPS. Note that our design can be readily generalized to type-I and type-II SPDC, as illustrated in Figure~\ref{fig:type12}. The PGR can be further enhanced with PPLN waveguides on the lithium niobate on insulator~(LNOI) platform. However, to fully unlock the performance potential of this chip-scale device, the facet coupling loss must be reduced. Furthermore, we show that our non-degenerated PEPS is able to faithfully teleport photonic states at different wavelengths by implementing teleportation. Our work provides a high-quality non-degenerated PEPS, which is beneficial for quantum networks that require the establishment of photonic entanglement across different wavelengths.

\begin{table*}[htbp]
\caption{A summary of non-degenerated sources. --: Cannot be inferred from the reported data.}\label{Tb:spdc}
\begin{threeparttable}
\begin{tabular}{p{30mm}p{26mm}p{30mm}p{18mm}p{25mm}p{15mm}p{10mm}}
    \hline\hline
    \multirow{2}{=}{Reference} &\multirow{1}{=}{Material} & \multirow{1}{=}{PGR/Coincidence$^\text{a}$} & \multirow{1}{=}{$\lambda_i$, $\lambda_s$} &\multirow{2}{=}{Source Setting$^\text{b}$} &\multirow{2}{=}{Fidelity} & \multirow{2}{=}{$\eta^\text{c}$} \\ 
    & \multirow{1}{=}{PM Type} &\multirow{1}{=}{[pairs/s/mW]} &\multirow{1}{=}{[nm]} & & \\
     \hline
    \multirow{2}{=}{\textbf{This work}}
    & PPLN
    & \multirow{2}{=}{$8.2\times10^5$/$3\times10^4$}
    & \multirow{2}{=}{810/1550}
    & SC-DP
    & \multirow{2}{=}{98.5\%}
    & \multirow{2}{=}{19.1\%}  \\
    & \multirow{1}{=}{Type-0}  
    & & 
    & Sagnac  \\
    \hline
    \multirow{2}{=}{Pelton~\emph{et al.}~\cite{Pelton2004OE}}
    & PPKTP
    & \multirow{2}{=}{$1.9\times10^5$/$1.5\times10^2$} 
    & \multirow{2}{=}{810/1550}
    & \multirow{2}{=}{DC-SP}
    & \multirow{2}{=}{89.85\%}
    & \multirow{2}{=}{2.8\%} \\
    & Type-I\\
    \hline  
    \multirow{2}{=}{K\"onig~\emph{et al.}~\cite{Konig2005PRA}}
    & PPLN
    & \multirow{2}{=}{$1.5\times10^4$/$4.5\times10^2$}
    & \multirow{2}{=}{795/1609}
    & SC-DP
    & \multirow{2}{=}{98.23\%}
    & \multirow{2}{=}{17.3\%}  \\
    & Type-I  
    & & 
    & MZI  \\
    \hline  
    \multirow{2}{=}{H\"ubel~\emph{et al.}~\cite{Hubel2007OE}}
    & PPKTP
    & \multirow{2}{=}{--/$1.3\times10^3$}
    & \multirow{2}{=}{810/1550}
    & \multirow{2}{=}{DC-SP}
    & \multirow{2}{=}{94.5\%}
    & \multirow{2}{=}{--}  \\
    & Type-I\\
    \hline  
    \multirow{2}{=}{Chatellus~\emph{et al.}~\cite{deChatellus2006OE}}
    & PPLN
    & \multirow{2}{=}{$(8.7\times10^4)^\text{d}$/--}
    & \multirow{2}{=}{810/1550}
    & \multirow{2}{=}{SC-SP}
    & \multirow{2}{=}{92.5\%}
    & \multirow{2}{=}{--}  \\
    & Type-I\\
    \hline
    \multirow{2}{=}{Sauge~\emph{et al.}~\cite{Sauge2008OE}}
    & PPLN
    & \multirow{2}{=}{$(3\times10^5)^\text{e}$/$3.7\times10^4$}
    & \multirow{2}{=}{810/1550}
    & SC-DP
    & \multirow{2}{=}{95.5\%}
    & \multirow{2}{=}{$35.1\%^\text{f}$}  \\
    & Type-I  
    & & 
    & MZI  \\
    \hline
    \multirow{2}{=}{Fiorentino~\emph{et al.}~\cite{Fiorentino2008OE}}
    & PPKTP
    & \multirow{2}{=}{--/$2.5\times10^3$}
    & \multirow{2}{=}{586/1310}
    & SC-DP
    & \multirow{2}{=}{90.5\%}
    & \multirow{2}{=}{--}  \\
    & Type-II
    & & 
    & MZI  \\
    \hline
    \multirow{2}{=}{Hentschel~\emph{et al.}~\cite{Hentschel2009OE}}
    & PPKTP
    & \multirow{2}{=}{$2\times10^5$/$9.3\times10^3$}
    & \multirow{2}{=}{810/1550}
    & SC-DP
    & \multirow{2}{=}{98.2\%}
    & \multirow{2}{=}{21.6\%} \\
    & Type-I  
    & & 
    & Sagnac  \\
    \hline
    \multirow{2}{=}{Clausen~\emph{et al.}~\cite{Clausen2014NJP}}
    & w-PPKTP/PPLN
    & \multirow{2}{=}{$10^5$/800}
    & \multirow{2}{=}{883/1338}
    & DC-SP
    & \multirow{2}{=}{98.05\%}
    & \multirow{2}{=}{9\%} \\
    & Type-I  
    & & 
    & MZI  \\
    \hline
    \multirow{2}{=}{Stuart~\emph{et al.}~\cite{Stuart2013PRA}}
    & PPLN
    & \multirow{2}{=}{$(4\times10^5)^\text{d}$/250}
    & \multirow{2}{=}{810/1550}
    & DC-SP
    & \multirow{2}{=}{97.4\%}
    & \multirow{2}{=}{2.5\%}  \\
    & Type-0  
    & & 
    & Sagnac  \\
    \hline    
    \multirow{2}{=}{Dietz~\emph{et al.}~\cite{Dietz2016APB}}
    & PPLN
    & \multirow{2}{=}{$5.8\times10^6$/$2.3\times10^4$}
    & \multirow{2}{=}{894.3/1313.1}
    & SC-DP
    & \multirow{2}{=}{75\%}
    & \multirow{2}{=}{6.3\%}  \\
    & Type-0  
    & & 
    & sandwich\\
    \hline
    \multirow{2}{=}{Aizawa~\emph{et al.}~\cite{Aizawa2023AO}}
    & PPKTP
    & \multirow{2}{=}{--/2}
    & \multirow{2}{=}{606/1550}
    & SC-DP
    & \multirow{2}{=}{94.4\%}
    & \multirow{2}{=}{--}  \\
    & Type-0  
    & & 
    & Sagnac\\
    \hline
    \multirow{2}{=}{Lee~\emph{et al.}~\cite{Lee2021QST}}
    & PMF~(SFWM)
    & \multirow{2}{=}{$1.2\times10^4$/90}
    & \multirow{2}{=}{764/1221}
    & SC-DP
    & \multirow{2}{=}{96.8\%}
    & \multirow{2}{=}{8.7\%}  \\
    & Type-0  
    & & 
    & MZI+Sagnac  \\
    \hline
    \multirow{2}{=}{Szlachetka~\emph{et al.}~\cite{Szlachetka2023APL}}
    & PPLN
    & \multirow{2}{=}{$6.17\times10^6$/$6.96\times10^4$}
    & \multirow{2}{=}{785/1651}
    & SC-DP
    & \multirow{2}{=}{96.72\%}
    & \multirow{2}{=}{10.6\%}  \\
    & type-0  
    & & 
    & Sagnac  \\
    \hline
    \multirow{2}{=}{Hong~\emph{et al.}~\cite{Hong2024OL}}
    & PPKTP
    & \multirow{2}{=}{--/--}
    & \multirow{2}{=}{548.2/1550}
    & SC-DP
    & \multirow{2}{=}{97.5\%}
    & \multirow{2}{=}{--}  \\
    & type-0  
    & & 
    & MZI  \\
    \hline
    \hline
\end{tabular} 
    \begin{tablenotes}
    \footnotesize
    \item[a] The photon pair generation rate~(PGR) is calculated by $\frac{N_s\times N_i}{C}$, where $N_s$, $N_i$ and $C$ represent the count rate of signal, idler and coincidence.  
    \item[b] DC-SP: double crystal-single pass; SC-DP: single crystal-double pass; sc-sp: single crystal-single pass; MZI: Mach-Zehnder interferometer.
    \item[c] The heralding efficiency is generally different for signal and idler photons. For the sake of convenience in comparison, we calculate $\eta$ by $\sqrt{\frac{C}{\text{PGR}}}$ under the assumption of $\eta_s=\eta_i$. 
    \item[d] The PGR is calculated by assuming $N_s=N_i$. 
    \item[e] The reported PGR includes coupling loss from free space to SMF. 
    \item[f] The heralding efficiency does not include coupling loss from free space to SMF.
    \end{tablenotes}
    \end{threeparttable}
\end{table*}

\bibliography{NDPS}
\end{document}